# Impact of Triaxiality on the Emission and Absorption of Neutrons and Gamma Rays in Heavy Nuclei

Eckart Grosse[a,b]*, Arnd R. Junghans[a], Ralph Massarczyk[a,b]

[a] *Institut für Strahlenphysik, Helmholtz-Zentrum Dresden-Rossendorf, Bautzner Landstraße 400, 01328 Dresden, Germany*
[b] *Institut für Kern- und Teilchenphysik der TU Dresden, Zellescher Weg 19, 01069 Dresden, Germany*

**Abstract:** For many spin-0 target nuclei neutron capture measurements yield information on level densities at the neutron separation energy. Also the average photon width has been determined from capture data as well as Maxwellian average cross sections for the energy range of unresolved resonances. Thus it is challenging to use this data set for a test of phenomenological prescriptions for the prediction of radiative processes. An important ingredient for respective calculations is the photon strength function for which a parameterization was proposed using a fit to giant dipole resonance shapes on the basis of theoretically determined ground state deformations including triaxiality. Deviations from spherical and axial symmetry also influence level densities and it is suggested to use a combined parameterization for both, level density and photon strength. The formulae presented give a good description of the data for low spin capture into 124 nuclei with $72 < A < 244$ and only very few global parameters have to be adjusted when the predetermined information on ground state shapes of the nuclei involved is accounted for.

## 1. Introduction

Since decades the shapes of heavy nuclei have been under intense investigation – experimentally, as reviewed *e.g.* in 2001 by Raman *et al.* as well as in theoretical studies as presented *e.g.* by Bohr and Mottelson in 1975. A good knowledge of nuclear shapes – including their dependence on energy and angular momentum – is of great importance for the understanding of nuclear fission and for respective model calculations. The experimental information on enhanced electric quadrupole (E2) transitions or moments indicates the presence of static quadrupole moments and thus the breaking of spherical symmetry in quasi all heavy nuclei away from magic shells. It was shown by Ericson in 1960 that nuclear deformation influences nuclear level densities and a data compilation presented in 1988 by Dietrich and Berman reveals its importance for the energy dependence of photon absorption in the region of the isovector giant dipole resonance IVGDR, which plays an important role for the extraction of electric dipole strength (E1). For low lying states of heavy nuclei also the breaking of axial symmetry has to be admitted, as shown by Cline, 1986, resulting from an analysis of heavy ion induced multiple Coulomb excitation based on rotation invariants as proposed by Kumar, 1972. Girod and Grammaticos, 1982, have pointed out the importance of triaxiality in calculations of fission barrier heights.

---

* Corresponding author.
*E-mail address:* e.grosse@tu-dresden.de.

Photonuclear reactions and other radiative processes influence schemes for the use of fission energy in everyday life – including the possibility to induce an accelerated decay of the waste produced in presently used fission reactors – and they have played a role in the cosmic element production. It thus seems justified to investigate the possible influence of the details of nuclear shapes including their triaxiality on the extraction of photon strength functions from IVGDR data as well as on nuclear level densities defining the final state phase space of nuclear reactions. The study presented here investigates 124 nuclides reached by neutron capture in spin-0 targets. It makes use of a Hartree-Fock-Bogoliubov (HFB) calculation presented in 2010 by Delaroche *et al.,* which predicts not only E2-transition strengths to the first $2^+$-states and hereby quadrupole deformation reasonably well, but also the breaking of axial symmetry, *i.e.* the triaxiality parameter $\gamma$. These predictions, made for the even nuclei between the proton and neutron drip lines, can be used as basis for a global survey on radiative cross sections if introduced into parameterizations for the energy dependence of photon strengths as well as of nuclear level densities. Predictions made accordingly will be compared to average radiative widths at the neutron binding energy $S_n$, and of capture cross sections in the energy range of 30 keV, which are of special interest for simulations concerning the element production in giant stars.

## 2. Photon strength
*2.1 Photon absorption in the isovector giant resonance IVGDR.*

The splitting of the IVGDR in the lanthanide and actinide nuclei is obvious from the experimental photo-neutron data compiled by Dietrich and Berman in 1988: The extracted parameters of the Lorentz curves as fitted to the seemingly double humped total cross sections are consistent with the notion that these are deformed, *i.e.* non-spherical. Apparently a new analysis as performed by Plujko *et al.*, 2011, including new IVGDR data, closely follows the spirit of the older work and also considers only these nuclides – and about 10 additional ones with A >72 – to have no spherical symmetry. This is at variance to the findings of Raman *et al.*, 2001, where it was shown that enhanced E2 transitions in nearly all heavy nuclei indicate a breaking of spherical symmetry. Junghans et al., 2008 and 2011, pointed out that the IVGDR shapes resembling a single Lorentzian curve should not be misinterpreted as a signature of sphericity but rather be analysed as a superposition of three such curves being separated by less than their widths. In this work it was shown that a small deviation from sphericity is likely to be accompanied by triaxiality, whereas in well deformed nuclei a small deviation from axial symmetry is expected – in accordance to the observations of Cline, 1986, Wu *et al.*, 1996 and of Andrejtscheff and Petkov, 1993. It was also shown, that a large number of the IVGDR data as compiled in the EXFOR data base are well described by a triple Lorentzian (TLO) given by

$$\sigma_{abs}^{E1}(E_\gamma) = C \frac{Z \cdot N}{A} \frac{2}{3\pi} \sum_{k=1}^{3} \frac{E_\gamma^2 \Gamma_k}{(E_\gamma^2 - E_k^2)^2 + E_\gamma^2 \Gamma_k^2}; \quad C = 2\pi^2 \frac{\alpha \hbar^2}{m_N} = 5.97; \quad \Gamma_k = 0.045 \cdot E_k^{1.6} \qquad (1).$$

The widths $\Gamma_k$ of the three IVGDR components at the energies $E_k$ can be well parameterized by the expression of Eq. 1. Generalizing a suggestion of Bush and Alhassid, 1991, originally formulated for the three components of one nucleus, the spreading width for all nuclei with A > 72 is parameterized in Eq. 1 with the exponent 1.6 derived from hydrodynamical considerations and the proportionality factor obtained from a fit to more than 20 different nuclei. Junghans *et al.*, 2008 and 2011, have also shown that Eq. 1 agrees to data not only in the range of the IVGDR, but also below, where the photon absorption excitation functions were determined *e.g.* by photon scattering. The situation in two more nuclei is shown in Figs. 1 and 2, for which Plujko *et al.*, 2011, have assumed spherical symmetry and extracted considerably larger widths as compared to Eq.1. This leads to an enhanced cross section at low $E_\gamma$, where the parameterization as presented here apparently requires extra strength as depicted by the thin lines, which will be discussed in Ch. 2.4.

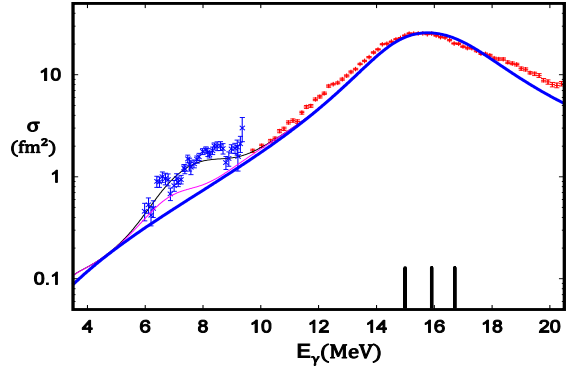

**Fig.1** presents data for the IVGDR in $^{118}$Sn. Cross sections for ($\gamma$,n) taken from EXFOR and from photon scattering as obtained by Axel et al., 1970, are compared to a TLO calculation shown as blue line. The magenta line indicates additional E1 strength and the black line results from also adding M1 strength. The black bars indicate the three pole positions corresponding to the triaxiality at low $E_x$.

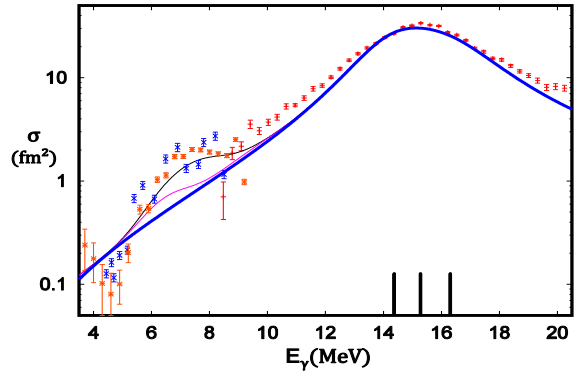

**Fig.2** depicts the data for the IVGDR in $^{138}$Ba. Cross sections for ($\gamma$,n) taken from EXFOR and from a scattering experiment with quasi-monochromatic photons (Tonchev et al., 2010, blue ×) are compared to a TLO calculation (blue line, cf. Fig. 1). Data from a bremsstrahlung experiment on $^{136}$Ba (red ×, Massarczyk et al., 2012) are shown for comparison. The black bars indicate the triaxiality.

The fit of the centroid energies $E_0$ resulted in an invariant mass $m_{eff}$ = 855 MeV with liquid drop model parameters determined by fitting ground state masses as proposed by Myers *et al.*, 1977. The factor Z·N/A in Eq. 1 corresponds to the TRK sum rule first proposed by Kuhn in 1925. As outlined in Beyer *et al.*, 2011, additional E1 strength as discussed by Gell-Mann *et al.*, 1954, is mainly concentrated at higher energy. As the IVGDR dipole oscillation is fast as compared to quadrupole modes the splitting can be treated adiabatically by inserting mean deformation parameters $\beta_2$ and $\gamma$ in the expression proposed by Hill and Wheeler, 1953, for the axis lengths of a triaxial body and for the inversely proportional energies.

$$E_k = E_0 \cdot \exp\left[-\sqrt{\frac{5}{4\pi}} \cdot \beta_2 \cdot \cos(\gamma - \frac{2}{3}k\pi)\right] \qquad (2)$$

Whereas from experimental data the deformation $\beta_2$ can be deduced for many nuclei near β-stability the triaxiality parameter $\gamma$ is known for a limited number of nuclides only. This is why Grosse *et al.*, 2012, indicated the possibility to use the $\beta_2$ and $\gamma$ as predicted by in 2010 by Delaroche *et al.;* further work along these lines is in progress. Obviously the TLO approach considers a triaxial nucleus to be the general case with the limits of vanishing $\gamma$ in the quasi-axial case of large $\beta_2$, and of $\beta_2$ tending to zero near magic shells (with $\gamma$ becoming meaningless). As Bertsch *et al.* have pointed out in 2007 that the HFB calculations tend to overestimate $\beta_2$ for the rare case of near magic nuclei; the axial deformation was reduced by a factor of 0.44 for nuclei with magic neutron or proton numbers. Thus the combination of the HFB predictions to the TLO parameterization is global and allows extrapolations away from the valley of stability.

*2.2 Electric dipole strength in and below the IVGDR*

The multipole strength functions $f_\lambda(E_\gamma)$ are related to the average photon absorption cross section in a given energy interval $\Delta_E$ by:

$$f_\lambda = \frac{<\sigma^\lambda_{abs}(E_\gamma)>}{g_{eff}(\pi\hbar c)^2 \overline{E}_\gamma^{2\lambda-1}} = \frac{1}{g_{eff}(\pi\hbar c)^2 \overline{E}_\gamma^{2\lambda-1} \Delta_E} \int_{\Delta_E} \sigma^\lambda_{abs} dE \quad (3)$$

The strength functions $f_\lambda(E_\gamma)$ are supposed to be direction independent and they are thus used for excitation as well as decay processes relating photon scattering to radiative capture and photonuclear processes; see *e.g.* Bartholomew *et al.*, 1972. Using that $f_\lambda$ is directly related to the electromagnetic decay widths of the resonant levels R in the interval Δ one gets:

$$f_\lambda = \frac{1}{\Delta_E} \sum_{R\in\Delta} \frac{\Gamma_{R\gamma}}{E_\gamma^{2\lambda+1}} \cong \frac{<\Gamma_{R\gamma}(E_\gamma)>}{D_R \overline{E}_\gamma^{2\lambda+1}} \quad (4).$$

The quantum-mechanical weight factor $g_{eff}$ will be discussed in the next chapter. $D_R$ denotes the average level spacing at the upper of the two levels connected by $E_\gamma = E_R - E_f$ and its insertion assures that for constant $f_\lambda$ the average resonance widths $<\Gamma_{R\gamma}(E_\gamma)>$ decrease with increasing level density $\rho_R = 1/D_R$. Often such decay takes place between levels which are both excited states and there is no simple way to study them starting from target ground states. But the average quantity $f_\lambda$ is insensitive to details of the nuclear spectrum and it thus makes sense to approximate any electromagnetic transition strength of energy $E_\gamma$ by $f_\lambda(E_\gamma)$ to be independent of the energies $E_R$ and $E_f$; this assumption first made by Brink in 1955 and published by Axel, 1962, is usually named Axel-Brink hypothesis.

As documented by EXFOR, photo-absorption and photo-dissociation data in the IVGDR region do not exist for all stable isotopes and photon strength results for energies below are even rarer. Results for energies below the neutron binding energy $S_n$ were partly taken from photon scattering *e.g.* using the bremsstrahlung facility at the Dresden radiation source ELBE or the Duke laser backscattering set-up HIγS described by Tonchev *et al.* in 2010. Such data complement the information from photo-dissociation studies and a wide energy range is covered leading to improved predictions for the region below 7 MeV, which has special importance for radiative neutron capture calculations, as will be outlined in Ch. 4. When the photon absorption cross section is derived from elastic photon scattering one exploits the fact that at sufficiently large angles nuclear resonance fluorescence is much stronger than Thomson and Delbrück scattering by the nuclear charge. As in such a compound nucleus like process the compound resonances may also decay in inelastic channels a correction method was developed on the basis of statistical considerations by Schramm *et al.*, 2011. It is based on the fact, that the electromagnetic strength is responsible for the absorption as well as the emission of photons, such that an iterative procedure can result in a self-consistent solution. A similar concept was already formulated previously by Axel *et al.*, 1970.

As depicted in Figs. 1 and 2 TLO also agrees to experimental data at low energies. This supports the sole dependence of $\Gamma_\gamma$ on the pole energies $E_k$ – which is at variance to a scheme with $\Gamma_\gamma$ depending on the photon energy $E_\gamma$ and on the excitation energy $E_x$ as proposed by Kopecky and Uhl, 1990. This scheme introduces additional parameters to bring the low energy strength based on the single Lorentzian into an agreement to data below $S_n$. But such fits are specific for each nuclide and can thus not serve as basis for a global parameterization – similar to the ones as presented by Plujko *et al.*, 2011.

*2.3. Special considerations for odd nuclei*

For even nuclei with spin $J_0 = 0$ in the ground state and $J_r$ in the excited level the $g_{eff}$ in Eq. 3 are identical to the quantum-mechanical weight factor, whereas for $J_0 \neq 0$ two facts have to be considered to assure the equality of the $f_\lambda$ as used in Eqs. 3 and 4: Photon absorption into a mode $\lambda$ populates m members of a multiplet with m = min($2\lambda+1$, $2J_0+1$) and the decay widths to the ground state $\Gamma_{0r}$ are equal for each member of the multiplet. The observed strength corresponds to the cross section summed over the multiplet and this can be described by an effective spin factor:

$$g_{eff} = g = \frac{2J_r+1}{2J_0+1} = 2J_r+1 = 2\lambda+1 \text{ for spin0 nuclei}; \quad g_{eff} = \sum_{r=1,m} \frac{2J_r+1}{2J_0+1} = 2\lambda+1 \text{ for odd nuclei} \quad (5).$$

Both conditions were shown by Bartholomew *et al.*, 1972, to be fulfilled in many heavy nuclei; they relate to the assumption of weak coupling between the odd particle and the mode $\lambda$. In the case of scattering by a target with non-zero ground state spin $J_0$ the observed strength corresponds to the cross section summed over a multiplet as described with Eq. 5 and the statistical factor which would have to appear is $2\lambda+1$; in such nuclei the IVGDR is a triplet corresponding to $\lambda=1$ (or a doublet for $J_0 = \frac{1}{2}$). The TLO-calculations for odd-A nuclei as shown in Fig. 3 were performed on the basis of these considerations and they agree to the experimental data similarly well as is the case for even nuclei, as seen in Fig. 3 showing a comparison of data for $^{88}$Sr and $^{89}$Y in the IVGDR as well as below $S_n$. The similarity of the data in both regions is obvious. Here as well as in the other spin nonzero nuclei deformations and radii were obtained by averaging the respective predictions for the even neighbours by Delaroche *et al.*, 2010; complementary calculations for odd nuclei are of interest here.

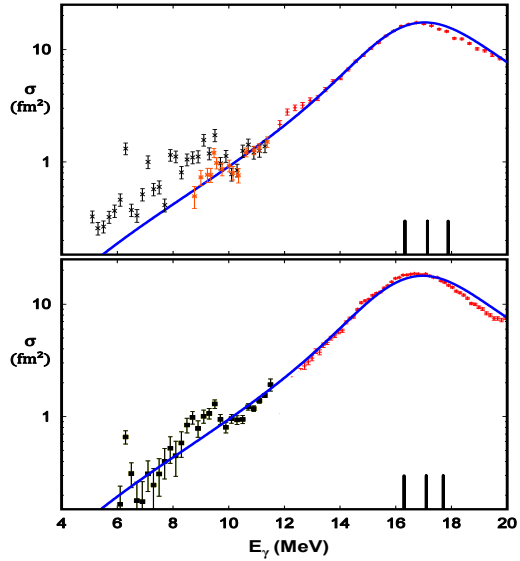

**Fig 3**: *TLO predictions (blue continuous curve) for the IVGDR in $^{88}$Sr (top) and $^{89}$Y, the poles of which are indicated in black. The measured cross sections of photo-neutron production taken from EXFOR are shown in red (+). Photon absorption as derived from scattering data is depicted in black (✕, Schwengner et al., 2007), (▯, Benouaret et al., 2009) and orange (✳, Datta et al., 1973). The TLO-calculations for $^{89}$Y were performed on the basis of the considerations related to Eq.5 and they agree to the experimental data similarly well as is the case of $^{88}$Sr.*

*2.4 Photon strength of other character than isovector electric dipole*

The height of the low energy tail, of special importance for predictions of the average radiative width and thus for the radiative capture cross section, is nearly proportional to the IVGDR width and does not strongly depend on its deformation induced splitting. But this splitting plays an important role in the adjustment of the width, if it is done by a fit to the data for a single nucleus. Such local fits performed with only one Lorentzian instead of using a close triplet result in a rather high dipole strength tail at small energy. In an article on the

Parameter Library RIPL-3 by Capote *et al.*, 2009, an explicit photon energy dependence of the IVGDR damping width was introduced as modified Lorentzian (MLO). Such a modification may result in a description of the data in the low energy tail without accounting for strength of other character than isovector electric dipole. The literature on photon scattering in this energy range has recently be reviewed by Grosse and Junghans, 2013, and the following components have been found to be considered:

      a. Isoscalar electric dipole strength in 'pygmy' resonances;
      b. Electric dipole strength resulting from coupling low energy E2 and E3 modes;
      c. Magnetic spin flip strength;
      d. Collective orbital magnetic strength (scissors mode, strong in deformed nuclei);
      e. Very low energy excitations (mainly M1, if resulting from orbital rearrangement).

Fragmented parts of the electric dipole strength outside of the IVGDR have attracted much interest, and also for magnetic modes interesting results were reviewed recently by Heyde *et al.*, 2010. High resolution photon scattering data as *e.g.* published by Endres *et al.*, 2009, show spectral details related to nuclear structure effects, which are especially significant in near magic nuclei; weak components are likely to be hidden in the experimental background. In the region of unresolved compound nucleus resonances a numerical averaging has to be applied in the derivation of the complete strength. In Figs.1 and 2 phenomenological approximations to data indicating the presence of strength not belonging to the IVGDR are shown as thin lines.

## 3. Level densities

Level densities in heavy nuclei can be calculated microscopically and in phenomenological models as listed by Capote *et al.*, 2009. The latter are usually based on thermo-dynamical considerations shown to be justified *e.g.* by Bohr and Mottelson, 1975. In a multi-particle system like a nucleus of sufficiently high temperature $t$ the logarithm of the state density $\ln(\rho(E_x))$ is mainly proportional to the nuclear entropy $S$. Ericson, 1960, has pointed out, that for Fermionic systems a critical temperature $t_c = 0.567 \cdot \Delta_0$ exists, below which a condensation into pairs, i.e. Bosons, reduces the entropy from $S = 2\tilde{a} \cdot t$ and changes the equation of state. The energy shift related to pairing is A-dependent and is usually quantified by $\Delta_0 = 12 \cdot A^{-1/2}$. The level density parameter $\tilde{a}$ is expected to rise linearly with the number of constituents A, but surface effects are predicted to result in a two component expression $\tilde{a} = \alpha \cdot (A + \beta \cdot A^{2/3})$. The theoretical prediction $\alpha = 1/16$ and $\beta = 2$ has been found as a kind of average result from various works in the review by Svirin, 2006. There also the impact of shell effects is discussed in terms of those already known from a liquid drop model fit to ground state masses. The respective table originating from Mengoni and Nakajima, 1994, is available from Capote et al., 2009. An energy dependent reduction of the shell correction $\delta W_o$ taken from there is introduced by a damping constant $\gamma = 0.64 \cdot A^{-1/3}$ and one gets $S = 2a \cdot t$ with

$$\frac{a}{\tilde{a}} = 1 + \delta W_o \frac{1 - e^{-\gamma a t^2}}{a t^2}; \quad E_x = a t^2 + E_b - n\Delta_0; \tag{6}$$

Ignatyuk *et al.* have shown in 1993 that the paired superfluid phase can be characterized with a smooth transition at $t_c$ and that $a_c$ and $E_c$ are obtained by inserting $t_c$ for $t$ in Eq. 6, which can be solved *e.g.* iteratively. As another effect of pairing the effective energy is corrected for a condensation energy $E_b = 1.5 \cdot a_c \cdot \Delta_0^2 / \pi^2$; by selecting n an additional difference between even (n=0) and odd (n=1) nuclei is introduced. Using Table 1 in Svirin, 2006, the entropy is given as a function of $t$, $t_c$ and $a_c$ as well as the determinant d resulting from the fixing of energy and nucleon numbers. In both phases the level density for small $J \ll \sigma$ is given by

$$\rho(E_x, J^\pi) \approx \frac{\chi \cdot (2J+1) \cdot e^S}{2\sigma^3 \sqrt{2\pi \cdot d}} \cdot \sqrt{\frac{\pi}{2}} \, \sigma_x \sigma_y \sigma_z; \quad \sigma_k = \sqrt{\frac{A m_N t}{5\hbar^2} (R_i^2 + R_j^2)} \tag{7}.$$

From *ad-hoc* considerations Svirin, 2006, derived the enhancement factor for rotational degrees of freedom for triaxial nuclei, which is the last term in Eq. 7; an additional correction factor χ is introduced to allow *a posteriori* an account for respective uncertainties. The product $\sigma_x \cdot \sigma_y \cdot \sigma_z$ effectively compensates the spin dispersion term $\sigma^3$ in the denominator in Eq. 7. All these parameters $\sigma_k$ are determined by the moments of inertia of the nucleus, which are assumed to be given by the axis ratios; rigid rotation and a sharp nuclear surface is assumed here. It should be pointed out, that the level density ρ in the 124 nuclei under study is determined by only one additional constant χ; the shape parameters from HFB-calculations as already discussed in relation to the IVGDR splitting have an influence on the $\sigma_k$ and thus on ρ of up to 30%. As the shape parameters are available from calculations extrapolations into domains away from stability can be based on this global parameterization. This clearly is an advantage over the fits applied usually, which minimize the deviations from data locally, *i.e.* for each nuclide separately. In Fig. 4 only one global parameter $\chi = 1/8 = 2^{-3}$ is applied which, according to Bohr and Mottelson, 1975, may be related to a level density reduction due to additional symmetry constraints in each of the 3 axes.

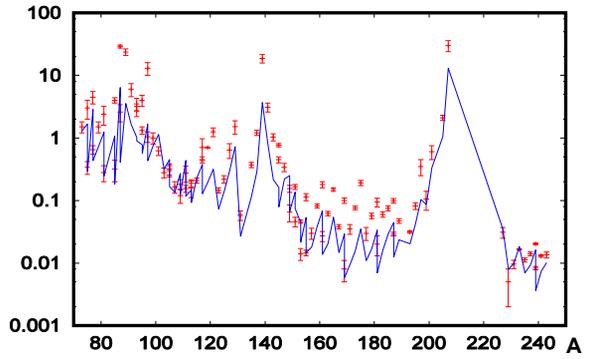

**Fig. 4**: *Average level distances 1/ρ in nuclei with 75<A<245 as observed in neutron capture by spin 0 target nuclei. Data (red +) compiled by Ignatyuk, 2006, are compared to the prediction presented in Ch. 3 (blue line).*

## 4. Radiative neutron capture
*4.1 Average photon widths*

The radiative capture of neutrons is of special interest to simulate isotope production *e.g.* in nuclear power plants and for network calculations of astrophysical processes. The good agreement of the low IVGDR energy slopes to our 'triple Lorentzian' parameterization (TLO) as obtained by using existing information on triaxial nuclear deformation suggests to use the corresponding photon strength function also for other electromagnetic processes like radiative neutron capture. To test the influence of dipole strength functions on radiative neutron capture over a wide range in A the investigation of only even-even target nuclei has the advantage of relying on a large sample with the same spin. For the s-wave capture by spin 0 nuclei the schematic approximation by Feshbach *et al*., 1947, and Hughes *et al*., 1953 yields the cross section:

$$\langle \sigma_R(n,\gamma) \rangle \cong 2(2\ell+1)\pi^2 \lambdabar_n^2 \rho(E_R, J^\pi=\tfrac{1}{2}^+) \cdot \langle \Gamma_{R\gamma}(E_\gamma) \rangle \qquad (8).$$

Here any $\ell$-dependent neutron strength enhancement is neglected. The effect of Porter-Thomas fluctuations in the region above separated resonances is reduced by averaging over a large number of neutron resonances R within $\Delta_R$; it was neglected. By replacing λ by 1 in Eq. 5 and summing over all final states $f \in \Delta_f = [0, S_n + E_R]$ and all resonances in $\Delta_R$ the average photon width is given by Bartholomew *et al*., 1972, to be:

$$\langle \Gamma_{R\gamma} \rangle = \sum_{f \in \Delta_f} \overline{\Gamma}_\gamma(R \to f) \cong \Delta_f \cdot \rho(E_f, J^\pi=\tfrac{1}{2}^+)\overline{\overline{\Gamma}}_\gamma(R \to f) \cong \int_{\Delta_f} M_t E_\gamma^3 f_1(E_\gamma) dE_\gamma; \quad M_t = \frac{\rho(E_f, J^\pi)}{\rho(E_R, J^\pi=\tfrac{1}{2}^+)} \cong 3 \cdot \frac{\rho(E_f)}{\rho(E_R)} \qquad (9)$$

The factor $M_t$ accounts for the number of magnetic sub-states reached by the γ-decay in comparison to the number of those populated by capturing the neutron and the double bar over $\Gamma_\gamma$ indicates the average in both intervals $\Delta_R$ and $\Delta_f$. As long as the validity of Eq. 9 is not limited by additional open channels it is obvious, that the slope of the capture cross section *vs.* $E_n$ is mainly determined by the neutron wave length. The absolute size of $\overline{\sigma_R}(n,\gamma)$ is proportional to the product of level density and photon strength such that the predicted cross section is most sensitive to $f_\lambda$ near $E_\lambda \approx$ 3-4 MeV. It was pointed out in 1990 by Kopecky and Uhl, that strength information can be extracted from capture data directly by regarding experimental average radiative widths, as shown in Fig. 5. Following Eq. 9 these are proportional to the photon strength and depend in addition only on the ratio $M_t$. In principle, $M_t$ as well as the average radiative width $\langle\Gamma(A,E_\gamma)\rangle$ would depend on a change in the spin distribution of $\rho(A,E_x)$ in the energy range reaching from $E_f$ to $E_R$. As indicated in Eq. 9, it is assumed here that for $\lambda = 1$-transitions from $J_R = 1/2^+$ to $J_f = 1/2^-$ and $J_f = 3/2^-$ the quantum-statistical part of $M_t$ is 3.

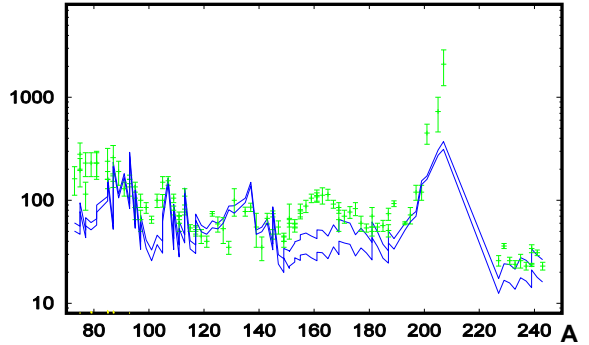

**Fig. 5:** *Average total radiative widths (green +) as compiled by Ignatyuk, 2006, in comparison to the prediction by Eq. 9 for the TLO photon strength (lower blue line). The upper line depicts the impact of the additional components listed in Ch. 2.4.*

*4.2 Maxwellian averaged cross sections*

A test on an absolute scale of level density predictions and of the photon strength as presented before can be obtained by applying Eqs. 8 and 9 to experiments on neutron capture. Concentrating on Maxwellian averaged cross sections (MACS) the following expression holds:

$$\langle\sigma(n,\gamma)\rangle_{kT} = \frac{2}{I_n\sqrt{\pi}}\int_0^{E_{max}}\sigma_R(E_n)E_n e^{-E_n/kT}dE_n = \frac{12}{I_n\sqrt{\pi}}\int_0^{E_{max}}(2\ell_n+1)\pi^2\lambdabar_n^2 E_n e^{-E_n/kT}\int_0^{E_R}\rho(E_f)E_\gamma^3 f_1(E_\gamma)dE_\gamma dE_n \quad (10)$$

$$I_n = \int_0^{E_{max}} E_n e^{-E_n/kT}dE_n; \quad E_{max} = 5kT; \quad T = T_{AGB} = 30\,\text{keV}; \quad E_R = S_n + E_n = E_\gamma + E_f$$

The folding of experimental cross sections as well as those given by Eq. 8 with a Maxwellian distribution of neutron energies is straightforward as pointed out by Käppeler *et al*, 2011. This work analyses nuclide abundance ratios and concludes that the relevant reactions are likely to happen at approximately kT = 30 keV. At such temperatures the capture rates are sufficiently small such that β-decays back to the valley of stability are at least as fast as the accretion of neutrons. In red giant (AGB) stars radiative neutron capture takes place starting at A ≈60 and a stepwise formation of heavier nuclei up to A ≈200 is possible by this slow "s-process". Data measured at respective neutron beams have been collected and tabulated as MACS by Dillmann *et al.,* 2010, covering many heavy nuclei. These data are free of effects resulting from strong resonances and thus can be considered a good representation of capture in the region of overlapping resonances. For several actinide nuclei equivalent data were compiled by Pritychenko *et al.*, 2010, and uncertainty bars were derived from their scatter. By only regarding the radiative capture by spin-zero targets effects related to ambiguities of spin cut off parameter and angular momentum coupling are suppressed, but still the data vary by about 4 orders of magnitude in the discussed range of A – and are well represented by the TLO-parameterization used here together with a schematic ansatz for $\rho(A, E_x)$, as is obvious from Fig. 6.

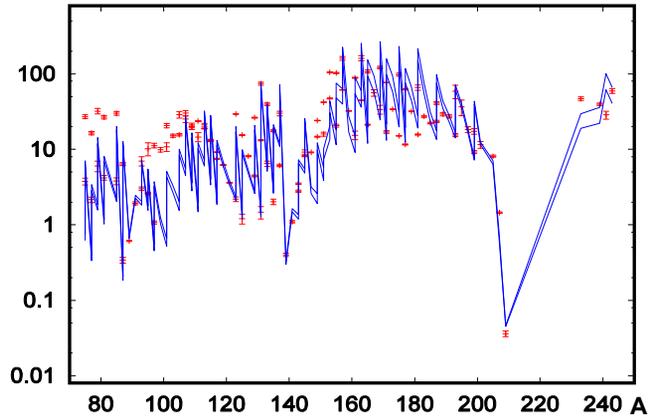

*Fig. 6: Maxwellian averaged cross sections as compiled by Dillmann et al., in 2010 (red +) in comparison to the prediction on the basis of Eq. 10 and the TLO photon strength (lower blue curve). The effect of other photon strength components is shown by the upper curve.*

## 5. Conclusions

   The splitting of the IVGDR in the nuclei with 72<A<244, for which respective data exist, is well described knowing the deformation and triaxiality from the calculations of Delaroche *et al.*, 2010. One parameter for the width and one for the centroid energies suffice for a satisfactory description of the resonance shape especially near the maximum of the cross section. There is no need for a strong deviation of the energy integrated strength from the classical dipole sum-rule (TRK). The widths vary smoothly with A and Z and are considerably smaller than those extracted in 1988 by Dietrich and Berman and by Plujko *et al.*, 2011. Thus a reduction of the damping width with photon energy is not needed to fit the photon strength data existing for the IVGDR tail in a number of nuclei. This results from the TLO description using three poles, and constitutes the main influence of the triaxiality on the photon strength determination. Using the literature study presented by Grosse and Junghans, 2013, on experimental photon scattering data it appears that magnetic and isoscalar electric dipole strength increases the radiative capture cross section by less than 25% only, such that the additional parameters for these small strength components are of reduced interest. The radiative neutron capture cross sections depend not only on the photon strength but in a similar amount also on the level density in the excitation region reached by the first photon emitted after capture. A good fit to experimental average radiative widths as well as to MACS of neutron capture by even nuclei results from the combination of the TLO photon strength and a scheme proposed by Svirin in 2006 to predict level densities in triaxial nuclei. Here one additional parameter $\chi$ has to be fixed by regarding data on level distances at $S_n$. An effective pairing energy shift and a shell correction energy are used which were taken from Mengoni and Nakajima, 1994, as available in RIPL-3 in 2014 for a large number of nuclei – as described by Capote *et al.* in 2009. Further investigations are needed concerning this point and others influencing level densities and their excitation energy dependence. Eventual anomalies in exotic nuclei as well as those near to closed shells need special attention.
   As shown in this paper, neutron capture cross sections in the energy range of overlapping resonances are well parameterized by the triple Lorentzian photon strength (TLO), when this is combined to a scheme for the prediction of level densities in nuclei of reduced symmetry, *e.g.* triaxial ones. The fact that for 124 spin-0 target nuclei with 72<A<244 the ratio of experiment over prediction is close to 1 albeit only a very small number of free parameters is needed – combined to quantities already determined independently – establishes the importance of triaxiality on photon strength and on radiative capture for neutrons. Thus this ansatz may be considered a very good starting point for network calculations in the field of nuclear astrophysics and especially for the element synthesis in the s- and p-processes. As it works up to actinide nuclei it is equally applicable for the numerical simulation of nuclear power systems and the transmutation of radioactive waste.

## Acknowledgements

Discussions with and support by K.H. Schmidt, R. Schwengner and A. Wagner are gratefully acknowledged.
## References

Andrejtscheff , W., Petkov, P. , 2001.  Phys. Lett. B 506, p. 239.
Axel, P., 1962.  Phys. Rev. 126, p. 671.
Axel et al. 1970. Phys. Rev. C 2, p. 689.
Bartholomew, G.A. et al. 1972. Adv. Nucl. Phys. 7, p. 229.
Benouaret, N. et al. 2009. Phys. Rev. C 79, p. 014303.
Bertsch, G.F. et al. 2007. Phys. Rev. Lett. 99, p. 032502.
Beyer, R. et al. 2011. Int. Journ. of Mod. Phys. E20, p. 431.
Bohr, A., Mottelson, B., 1975. Nuclear Structure, ch. 4 & 6, Benjamin Inc., Reading, Mass.
Brink, D., 1955. Ph.D. thesis, Oxford.
Bush, B., Alhassid, Y., 1991. Nucl. Phys. A 531, p. 27.
Capote , R. et al. 2009. Nucl. Data Sheets 110,  p. 3107; id., http://www-nds.iaea.org/RIPL-3/.
Cline, D.,1986. Ann. Rev. Nucl. Part. Sci. 36, p. 683.
Datta, S. et al.1973. Phys. Rev. C 8, p. 1421.
Delaroche, J-P. et al., 2010, Phys. Rev. C 81, p. 014303.
Dietrich, S.S., Berman, B.L., 1988. At. Data and  Nucl. Data Tables 38, p. 199.
Dillmann, I. et al., 2010. Phys. Rev. C 81, p. 015801; id., AIP Conf. Proc. 819, 123; http://www.kadonis.org/.
Endres, J. et al.,  2009. Phys. Rev. C 80, p. 034302; id., Phys. Rev. Lett. 105, p. 212503.
Ericson, T. 1960. Advances in Physics, 9, p. 425.
EXFOR  Database, 2013.  http://www.nndc.bnl.gov/exfor/exfor00.htm.
Feshbach, H.  et al., 1947. Phys. Rev. 71, p. 145.
Gell-Mann, M. et al., 1954. Phys. Rev. 95, p. 1612.
Girod,M., Grammaticos, B., 1982. Phys. Rev. C 27, p. 2317.
Grosse, E. et al., 2012. Eur. Phys. Journ., Web of Conf., 21, p. 04003; http://dx.doi.org/10.1051/epjconf/20122104003
Grosse, E., Junghans, A.R., 2013. Landolt-Börnstein, New Series I, 25, p. 4.
Heyde, K. et al.,  2010. Rev. Mod. Phys. 82, p. 2365.
Hill, D.L., Wheeler, J.A., 1953. Phys. Rev. 89, p. 1102
Hughes, D.J. et al., 1953. Phys. Rev. 91, p. 423.
Ignatyuk, A.V. et al., 1993. Phys. Rev. C 47, p. 1504.
Ignatyuk, A.V., 2006.  IAEA-TECDOC-1506 (2006); www-nds.iaea.org/RIPL-3/resonances.
Junghans, A.R. et al., 2008. Phys. Lett. B 670, p. 200.
Junghans, A.R. et al., 2011.  Journ. Korean Phys. Soc. 59, p. 1872.
Käppeler, F., 2011. Rev. Mod. Phys. 83, p.157.
Kopecky, J., Uhl, M., 1990. Phys. Rev. C 41, p. 1941.
Kuhn, W., 1925.  Z. f. Phys. 33, p. 408.
Kumar, K., 1972. Phys. Rev. Lett. 28, p.  249.
Massarczyk, R. et al., 2012. Phys. Rev.C 86, p. 014319.
Mengoni, A., Nakajima, Y., 1994. J. Nucl. Sci. Tech. 31, p. 151.
Myers, W.D. et al., 1977. Phys. Rev. C 15, p. 2032.
Plujko, V.A. et al., 2011. At. Data and Nucl. Data Tables 97, p. 567; http://www-nds.iaea.org/RIPL-3/gamma
Raman, S. et al., 2001. At. Data and Nucl. Data Tables 78, p. 1.
Schramm, G. et al., 2011. Phys. Rev. C 85, p. 014311
Schwengner, R. et al., 2007. Phys. Rev. C 76, p. 034321.
Svirin, M.I., 2006. Phys. of Part. and Nuclei, 37, p. 475.
Tonchev, A.P. et al., 2010. Phys. Rev. Lett. 104, p. 072501.
Wu, C.Y. et al., 1996. Nucl. Phys. A 607, p. 178; id., Phys. Rev. C 54, p. 2356.